\documentclass[onecolumn,11pt]{IEEEtran}
\usepackage{cite}
\usepackage{amsmath,amssymb,amsfonts}
\usepackage{algpseudocode}
\usepackage{algorithm}
\usepackage{graphicx}
\usepackage{textcomp}
\usepackage{enumitem}
\usepackage{xcolor}
\usepackage{caption}
\def\BibTeX{{\rm B\kern-.05em{\sc i\kern-.025em b}\kern-.08em
    T\kern-.1667em\lower.7ex\hbox{E}\kern-.125emX}}
\begin{document}
\title{An Enhanced Passkey Entry Protocol for Secure Simple Pairing in Bluetooth
\thanks{* Research supported by NSERC RGPIN 2016-05610.}}
\author{\IEEEauthorblockN{Sai Swaroop Madugula and Ruizhong Wei
}\\
\IEEEauthorblockA{\textit{Department of Computer Science} \\
\textit{Lakehead University}\\
Thunder Bay, Canada \\
{\tt rwei@lakeheadu.ca}
}
}
\maketitle

\begin{abstract}
Bluetooth devices are being used very extensively in today’s world. From simple wireless headsets to maintaining an entire home network, the Bluetooth technology is used everywhere. However, there are still vulnerabilities present in the pairing process of Bluetooth which lead to serious security issues resulting in data theft and manipulation. We scrutinized the passkey entry protocol in Secure Simple Pairing (SSP) in the Bluetooth standard v5.2. In this paper, we propose a simple enhancement for the passkey entry protocol in the authentication stage 1 of Secure Simple Pairing using preexisting cryptographic hash functions and random integer generation present in the protocol. The new protocol is more secure and efficient than previous known protocols. Our research mainly focuses on strengthening the passkey entry protocol and protecting the devices against passive eavesdropping and active Man-in-the-middle (MITM) attacks in both Bluetooth Basic Rate/Enhanced Data Rate (BR/EDR) and Bluetooth Low Energy (Bluetooth LE). This method can be used for any device which uses the passkey entry protocol.
\end{abstract}


\section{Introduction}
The Bluetooth is a simple wireless technology developed to exchange data over short ranges of area. The Bluetooth technology usually employs a wireless personal area network (PAN), also known as a piconet, in which data between two devices are exchanged securely. With the introduction of Bluetooth v4.0, namely Bluetooth Low Energy (Bluetooth LE), it became largely famous for its low power consumption. Bluetooth LE largely has its applications in the healthcare, fitness, security, and home network systems. This is because Bluetooth LE offered very low energy consumption while maintaining a similar communication range to that of Bluetooth Basic Rate/Enhanced Data Rate (BR/EDR). As the Bluetooth LE is known for its low energy consumption, it is most suited for the Internet of things (IoT) and is extensively used in it. The Bluetooth LE is ideal to send small amounts of data between devices continuously or periodically. Although different from each other, most security protocols in Bluetooth LE work like the BR/EDR. This is to ensure that both devices which support BR/EDR and Bluetooth LE are compatible to connect with each other. The BR/EDR and Bluetooth LE has a secure way of pairing and generating the link key which is essential for communication between any two Bluetooth devices. This method is known as Secure Simple Pairing (SSP).  The two main goals of SSP protocol are to protect the devices against passive eavesdropping and active attacks like MITM. 

SSP employs four association models that are used for authentication between two devices. One of the models  is based on the input-output compatibility (IOcap) of the two devices. In these models, the passkey entry protocol uses a 6-digit passkey which is shared on both sides by the user and is used as an authentication token. The Bluetooth standard describes that the passkey entry protocol provides protection against passive eavesdropping and active man-in-the-middle (MITM) attacks. However, the security of the protocol is based
on the assumption that the passkey is one-time.
There are vulnerabilities present in the passkey entry model which allows an attacker or an adversary to guess the passkey if it is reused and hijack the session which enables the attacker to gain access to the Bluetooth devices and retrieve sensitive information. 


The most recent countermeasure to the passkey reuse issue was proposed by Da-zhi Sun and Mu \cite{dazhimitm} in their improved passkey entry protocol. Their improved protocol successfully prevents an adversary from deducing the passkey using passive eavesdropping. In this paper, we have deduced that their protocol is still vulnerable to an attack algorithm, which is a variant 
of the normal brute-force method. By using our attack, the attacker will be able to deduce the correct passkey which will help them to conduct a successful MITM attack allowing them to gain access to the Long Term key (LTK).

\section{Problem Definitions}
\noindent To understand the vulnerability in the passkey entry protocol, we need to consider a few points.


\subsection{Security of passkey entry protocol according to Bluetooth standard}
The Bluetooth standard uses public key exchange to establish a common key between two devices. However, since
the public keys used in the protocol do not have certificates (they cannot use internet to verify the certificates), 
the key exchange is exposed to MITM attacks. 
According to the Bluetooth standard v5.2 \cite{sig}, the passkey entry protocol makes use of a 6-digit passkey which is entered in both
of the devices for authentication. This method is used to test if there is an MITM attack in the key exchange.
The 6-digit passkey is entered by the user and is of 20-bit length. The passkey entry protocol uses a method of gradual disclosure of each bit of the passkey based on the commitment protocol.  It describes that a simple brute-force guesser succeeds guessing the passkey with a probability of 0.000001 thus making it very difficult for an attacker to obtain the passkey. 
Then it  uses the Elliptic Curve Diffie Hellman (ECDH) under secure connections mode to protect the devices against passive eavesdropping. 

\subsection{Vulnerability in Passkey Entry Protocol}

The passkey entry protocol normally protects the devices against passive and active attacks. The Bluetooth standard requires that  the user should never reuse the passkey from the previous session.  However, if the passkey is reused, it creates a vulnerability which the attacker can exploit. This is caused by the method of verifying the passkey. If an attacker can passively capture the public key exchange packets and the passkey entry protocol communication packets, they can easily deduce the passkey and conduct a MITM attack during the user’s next SSP session.

\subsection{How does the vulnerability arise}
As we described earlier, ideally it is said that the user should not reuse the same passkey again. However, 
just like a user tends to use the same password for every account, the user might find it simple or easy to reuse the same passkey for every Bluetooth connection. The attacker can make use of this vulnerability and exploit the devices. 
Alternatively, the protocol which is generating the random passkey might reuse the same passkey to save space or its computational power which makes the passkey a static key. Since the random passkey generation protocol does not belong to the Bluetooth standard, it does not necessarily adhere to its rules. Therefore, there is a chance that it will use the same passkey for several connections.

\subsection{Assumptions on the capabilities for an adversary to exploit the vulnerability}
In this paper, we assume that an attacker has a set of capabilities that enables them to perform passive eavesdropping and active MITM attacks. The capabilities are defined below.

\begin{itemize}
\item The attacker has the knowledge of the frequency hopping pattern which is shared by the two legitimate Bluetooth devices.
\item The attacker has the ability to passively capture the transmission packets exchanged by the two legitimate Bluetooth devices.
\item The attacker has the ability to modify packets that are transmitted in real-time between the legitimate Bluetooth devices thus enabling them to perform a MITM attack before the devices successfully establish the passkey.
\end{itemize}

\subsection{Goal of an attacker}
The obtaining of the 6-digit passkey in the passkey entry protocol will enable an attacker to obtain the Long Term Key (LTK) and the encryption key which is derived after the secure simple pairing process is finished successfully. If an attacker can successfully conduct a MITM attack using the derived 6-digit passkey, then they can obtain the LTK and encryption key and will be considered as a trusted device by the legitimate devices. By using these keys, an attacker can simply make a request of sensitive data to the legitimate user. The legitimate user accepts the request since the connection was successful and sends sensitive data to the attacker thinking that it is an authentic device. Alternatively, the attacker can intercept the data being exchanged by legitimate devices and modify the data. Therefore, the vulnerability of the passkey entry protocol can lead to serious data manipulation and data theft. 

\subsection{Research Goal}

In this paper, we propose a new enhanced passkey entry protocol that provides protection against passive eavesdropping and MITM attacks even in case the user reuses the same passkey. Our method also tried significantly decreasing the communication cost and the computation cost. The main idea of our protocol is completely preventing an attacker from obtaining the correct passkey. By using our protocol, a legitimate device can successfully prevent an attacker from obtaining the LTK by MITM attacks.

\section{Association Models in SSP}

The SSP is the most important protocol in secure BLE pairing. The main goal of the SSP is to generate a shared Link key between two Bluetooth devices. A shared link key must be linked with a strong encryption algorithm to give the user protection against passive eavesdropping. Therefore, SSP uses Elliptic Curve Diffie Hellman (ECDH) to generate the Link Key (LK) and is used to maintain the pairing. However, The SSP process can still be subjected to man-in-the-middle (MITM) attacks because of the lack of user authentications.
 In order to provide authentication, the Bluetooth standard has introduced four associated models which are implemented in the authentication stage 1 (phase 2) of the SSP. The models are as follows (see \cite{sig}): 

\begin{description}

\item[$\bullet$ Just Works model:] 
This model is ideally used for scenarios where at least one device has no output display or any input capability, e.g. wireless speaker. The just works model does not provide any protection against MITM attacks. 
\item[$\bullet$ Numeric Comparison model:]
This model is used when both devices have display capabilities as well as input capability for the user to enter “yes” or “no”. This user confirmation is required for the device to know that the other device is legitimate and authentic and is, in fact, the same device the user is trying to connect to.  The numeric comparison model offers limited protection against MITM attacks with the attacker having a success probability of around 0.000001 on each iteration of the protocol.
\item[$\bullet$ Out of Band model:] 
This model is used when both devices can support communication over additional wireless channels e.g. near field communication (NFC). The out of band (OOB) channel ideally is resistant to MITM attacks. If not, there is a possibility that the data might be compromised during the authentication phase.
\item[$\bullet$ Passkey Entry model:] 
This model is used in scenarios where one device has input capability but does not have any output display capability. In this case, the user enters an identical 6-digit passkey on both devices. Alternatively, the passkey might be randomly generated by an algorithm in one device and displayed, which the user then inputs into the other device.
\end{description}

\section{Related Work on SSP}\label{AA}
Da-Zhi Sun and Li Sun \cite{dazhi} designed a formal security model to evaluate SSP’s association models and authenticated link key security. They discussed about the security and the possible vulnerabilities present in the SSP's association models. This model simulates the networking of the Bluetooth devices and detects attack vectors when an SSP session is run over an insecure public channel. Eli Biham and Lior Neumann \cite{eli} discovered a new vulnerability in the ECDH key exchange process. Their attack successfully allows an adversary to recover the session encryption which is used for secure data transfer with a success rate of 50\% during the pairing attempts. They accomplished this by modifying the y-coordinate of public keys of device and setting them to zero. By doing this, the computed DHkey value will always be infinity. Therefore, an adversary could calculate the encryption key this way. This attack is performed during the public key exchange phase of SSP. Samta Gajbhiya \cite{samta} proposed SSP with an enhanced security level which involves SSP with authenticated public key exchange and delayed-encrypted capability exchange. The main goal of their protocol is to prevent an attacker from forcing the legitimate devices to adopt the association model as per the attacker's convenience. Giwon Kwon \cite{giwon} introduced a method to increase the length of the Temporary Key (TK) by repeating it. They defined a model in which the master is required to select and transmit a security level through the security level field in the PairingRequest packet.  Da-Zhi Sun \cite{dazhimitm} proposed a novel method of protecting the passkey entry model in secure simple pairing protocol against MITM even if passkey is reused. After the passkey is injected into two devices, they generated a random nonce on both devices A and B and exchange it. Then, A computes a hash using the HMAC-SHA256 algorithm based on the passkey entered and sets the value \textbf{r$\boldsymbol{^*_a}$} by taking the six most significant digits of r. The same process is also done at device B and the rest of the protocol is kept the same. Due to the random nonces, the \textbf{r$\boldsymbol{^*}$} values computed from the hash are different for each connection. Barnickel \cite{barnickel} researched on the vulnerability on the passkey entry model of Bluetooth v2.1+ when the user reuses the same passkey for another SSP session. They implemented the MITM attack with a successful probability of 90\% using the GNU radio platform on Universal Software Radio Peripheral (USRP) and USRP2 devices. They proposed two countermeasures for the MITM attacks. The first is to record all the previously entered passkeys by the user and reject them if they are used again. However, this would mean there would be a very high storage cost of the protocol since it has to store every passkey entered for every SSP session. In addition to this, it is difficult for a user to enter a new password every time they want to connect to a new device since even they have to remember the previous passkeys used. The second countermeasure involved using encrypting and decrypting the random nonces which would prevent a passive attacker from eavesdropping. However, it greatly increases the complexity and the performance cost of the protocol due to the encryption and decryption functions.

\section{Architecture of Secure Simple Pairing}
For any kind of data transfer between two or more Bluetooth devices, they first need to establish a connection and pair. To make it secure, the devices use protocols such as Secure Simple Pairing under different security modes. The end goal of any pairing process between two Bluetooth devices is to generate a shared long-term link key on both sides which is used to transfer data. In classic Bluetooth and Bluetooth LE, there are a total of five phases in the SSP protocol. SSP strives to provide security while maintaining minimal user interaction. The primary focus of our research work is the passkey entry protocol present in the authentication stage 1 phase.

The SSP is the most important pairing protocol in Bluetooth Pairing. Before SSP, the Bluetooth technology used a 4-digit PIN code to pair with other devices. This PIN code was also a part of the Link Key calculation process. Due to the short length of the PIN code, several vulnerabilities were revealed, and to mitigate them, SSP was used. The end goal of the SSP is to generate a shared Link key between two Bluetooth devices.  However, The SSP process can still be subjected to man-in-the-middle (MITM) attacks. To prevent MITM attacks, the Bluetooth standard has introduced association models that are implemented in the authentication stage 1 (phase 2) of the SSP. 

\begin{figure}[h]
\centering\includegraphics[width=3 in]{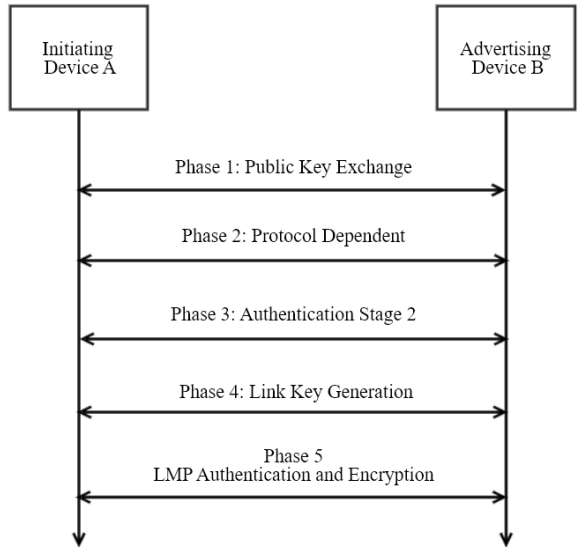}
\caption{SSP Architecture}
\label{Fig:SSPPhases}
\end{figure}

As shown in Fig.~\ref{Fig:SSPPhases}, the working of SSP is the same for any device, excluding phase 2 which is protocol dependent. All phases of SSP must work in the same way in every Bluetooth device. As for phase 2, when the IO capabilities are exchanged before phase 1 of SSP, both devices show what protocols are supported by them. Based on the compatibility and the requirements, a protocol is selected and executed.

\subsection{How SSP provides Security}
Usually, an attacker has two ways of obtaining data. The first technique is to execute passive eavesdropping (usually performed with specially designed tools like Ubertooth) and capture the transmission packets. By doing this, the attacker can try to compute the LTK based on the captured data in phase 1 and phase 2. If the attacker obtains the LTK in this way, then they can simply calculate the encryption key which is originally derived from the LTK. Thus, when two legitimate devices transfer data using this encryption key, the attacker can again capture the data packets and simply decrypt them using the deduced encryption. This way, they can gain access to sensitive data.

The second technique is to execute a MITM attack. This technique is more dangerous because if successful, the attacker has a trusted connection with the legitimate device and can request data or send malicious files to the target device. In the case of passive eavesdropping, the attacker can only access the data which is transferred between two legitimate devices. However, when it comes to active attacks such as MITM, the attacker and the target devices share the same LTK. Therefore, an attacker can enable data transfer requests whenever they require provided that the target device has its Bluetooth switched on.

 SSP provides protection from these attacks in the authentication stage 1 and authentication stage 2. In authentication stage 1, the SSP tries to prevent any MITM attacks by using either the numeric comparison or the passkey entry protocol based on the IOcap of devices. If an attacker was able to intercept the communication in phase 1 of SSP, it prevents the attacker from obtaining the LTK by performing authentication in phase 2. In authentication stage 2, the DHkey which is computed in phase 1 is verified which protects the device from passive eavesdropping. 

\subsection{Outline of SSP}

In phase 1 of SSP, the initiating device A and the advertising device B establish a common key DHkey following the Elliptic
Curve Diffie Hellman protocol. Note that since both devices do not have authentications, this key only can be used to prevent
eavesdropping attacks in the following, but not for the MITM attacks.

The SSP Phase 2 ( Authentication Stage 1)
 consists of 3 association models. In Bluetooth LE, the just works model integrates with the numeric comparison model. Aside from these two models, we have the out of band model and the passkey entry model. This phase is dependent on the IO capabilities of both devices. Based on the IOCap, the compatible association model is selected. Since our  focus is on the passkey entry model, we will not discuss the other association models in detail.

\begin{figure}[htbp]
\centerline{\includegraphics[width=0.6\textwidth]{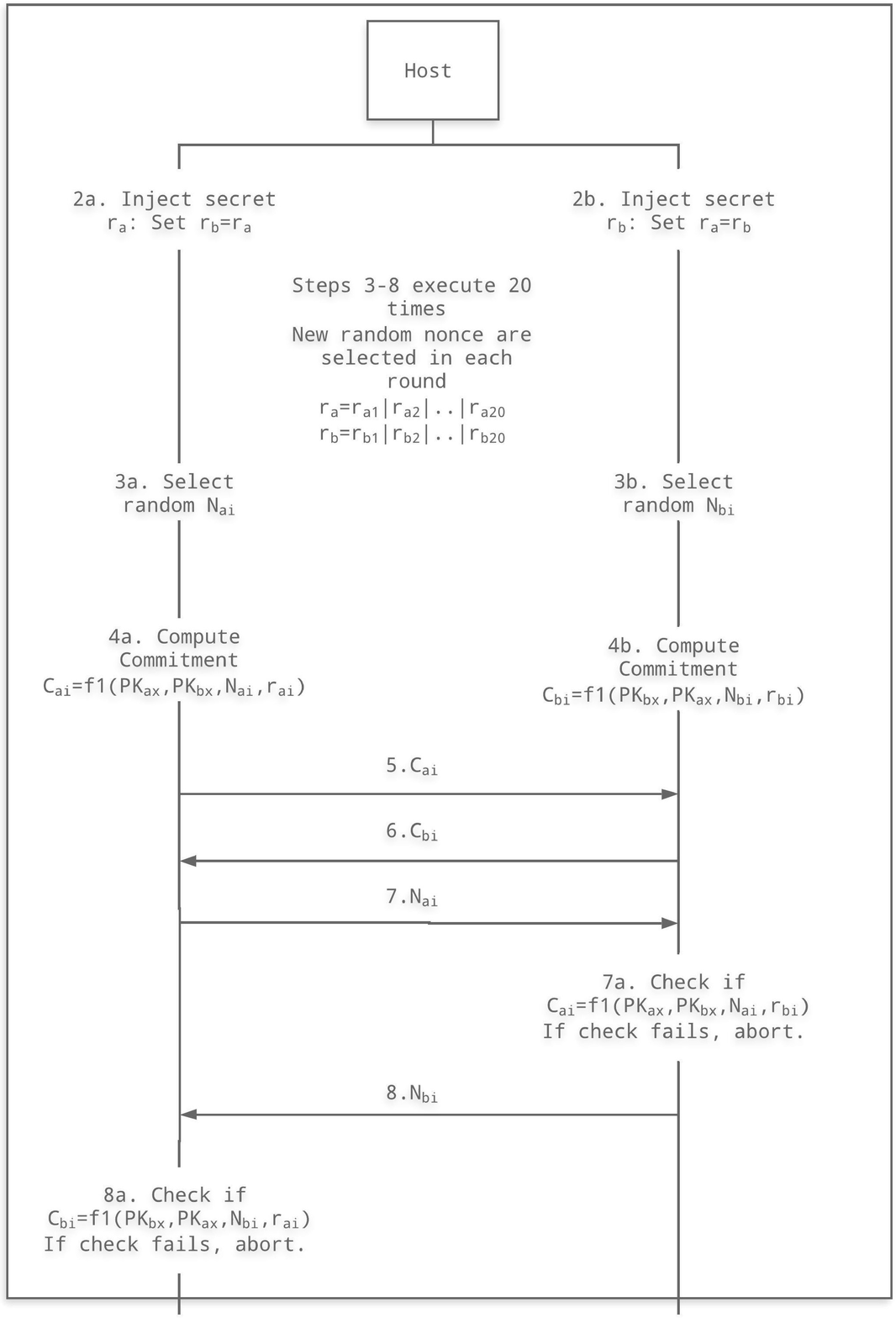}}
\caption{SSP Phase 2: Passkey Entry Protocol}
\label{Fig:SSP2}
\end{figure}

Figure \ref{Fig:SSP2} describes the passkey entry model.
In this model, the host or the user enters a 6-digit identical passkey on both devices A and B. The 6-digit passkey which is of 20-bit length is then stored into \textbf{r$\boldsymbol{_a}$} and \textbf{r$\boldsymbol{_b}$}. After injecting the passkey, the steps 3-8 are repeated for 20 times and each round is denoted with the integer $i$ which has a range of $1<=i<=20$. For every round, exactly one bit of passkey is sent to the other device. In each round, a 128-bit random nonce \textbf{N$\boldsymbol{_{ai}}$} and \textbf{N$\boldsymbol{_{bi}}$} is generated accordingly at both sides. After that, the device A computes a commitment value \textbf{C$\boldsymbol{_{ai}}$} using $f1$(PK$_{ax}$,PK$_{bx}$,N$_{ai}$,r$_{ai}$) and sends it to device B. B also generates the commitment value \textbf{C$\boldsymbol{_{bi}}$} using $f1$(PK$_{bx}$,PK$_{ax}$,N$_{bi}$,r$_{bi}$) and sends it to device A. After exchanging the commitment values, the device A sends its \textbf{N$\boldsymbol{_{ai}}$} value to B. Now, the device B, uses the \textbf{N$\boldsymbol{_{ai}}$} value sent by A to check if C$\boldsymbol{_{ai}}=f1$(PK$_{ax}$,PK$_{bx}$,N$_{ai}$,r$_{bi})$ and if the check passes, the device B will also send its \textbf{N$\boldsymbol{_{bi}}$} value. If the check does not pass, the protocol is aborted and the SSP session will need to be run again. The same procedure is followed at the device A also. According to the Bluetooth standard, the gradual disclosure of the passkey in this way prevents leakage of more than 1 bit of the passkey making it difficult for an MITM attacker to crack it. At the end of this stage, the values of \textbf{N$\boldsymbol{_{a}}$} and \textbf{N$\boldsymbol{_{b}}$} are set to \textbf{N$\boldsymbol{_{a}}20$} and \textbf{N$\boldsymbol{_{b}}20$} which are later used in the next stage, authentication stage 2. Here, $f1()$ is an HMAC-SHA256 
secure hash function used to calculate the commitment values of \textbf{C$\boldsymbol{_{ai}}$} and \textbf{C$\boldsymbol{_{bi}}$}. In the following $f2(), f3()$ are also HMAC-SHA256 based hash functions.

In SSP Phase 3 (Authentication Stage 2),
 the device A computes a new confirmation value Ea using $f3(DHkey, N_a, N_b, r_b, IOcapA, A, B)$ and B computes a confirmation value E$_b$ using $f3(DHkey, N_b, N_a, r_a, \\ IOcapB, B, A)$ 
 simultaneously. After the computations, the device A sends the E$_a$ to device B where B checks the E$_a$ with $f3(DHKey, N_a, N_b, r_b, IOcapA, A, B)$. If the check fails, the protocol aborts. If the check passes, the value of E$_b$ is sent to device A where A also checks the E$_b$ with $f3(DHKey, N_b, N_a, r_a, IOcapB, B, A)$. If the check fails, the protocol aborts. Here the value of \textbf{r$\boldsymbol{_a}$} and \textbf{r$\boldsymbol{_b}$} is the entire 20-bit passkey.

In SSP Phase 4,
 a shared link key is generated by using $f2(DHkey, N_a, N_b, btlk, BD\_ADDR_a, BD\_ADDR_b)$ at both sides. The order of the parameters must be the same in order to calculate the same key. The end goal of an attacker is to obtain this key.

Once the shared link key is established at both sides, the authentication process and the encryption key generation process is completed based on the link key. The steps followed in this phase are identical to the ones in the legacy pairing. Since we are only focusing on the passkey entry model, we omitted the  details of  phase 5 here.

\section{Vulnerability in Passkey Entry Protocol}

 The Bluetooth standard uses ECDH public key exchange in phase 1 to prevent passive eavesdropping. However,
 Since both devices do not have any authentication during the key exchange, an attacker can easily perform
 MITM attack which will result DHkeys shared with the attacker and the two devices. To prevent such attacks, the standard uses the Phase 2 or authentication stage 1 phase.  The main idea is that if there is a MITM attack in Phase 1, then the Bluetooth standard will try to prevent the attacker from connecting by using one of the association models in Phase 2. The  Passkey Entry model of Bluetooth  uses the commitment protocol to make sure that both parties have the same passkey. 
 That method works well if the user will change the passkey each time when the connection has a problem. Unfortunately, in practice very likely users will use the same passkey for many times, which will cause serious security problems.

For instance, a passive attacker can capture all the commitment values $(C_{ai}$ and $C_{bi})$, random nonce $(N_{ai}$ and $N_{bi})$ and deduce each bit of passkey {\tt r} by computing their own commitment $f1(Pk_{ax}, PK_{bx}, N_{ai}, r'_{ai})$ and comparing it with the original value. Since the value of the r$_{ai}$ is always either a 1 or 0, the attacker can easily obtain the entire passkey. In this way, the attacker can figure out the correct $r_{ai}$.
Ideally, we should not reuse the same passkey again, however, most users tend to reuse the old passkey again for simplicity and ease. This enables the attacker to launch an MITM attack with the known passkey successfully communicating with the two devices simultaneously.
In this way, after capturing all the packets and deducing the passkey, the attacker will successfully break into the system.

In what follows,  we  always assume that there is a MITM attack in Phase 1, and we what to check if the authentication process in Phase 2 can  always find out the attacks.

\section{An Improved Passkey Entry Protocol by  Sun and Mu}

Sun and Mu in \cite{dazhimitm} proposed an improved passkey entry protocol as a countermeasure to passkey reuse. We will refer to this protocol as the SM protocol. They proposed an improved passkey entry protocol in which,
 before  sending each bit of passkey (steps 3-8) in Figure~\ref{Fig:SSP2}, they generate a new random nonce on both sides and exchange them. Then, they compute $r=f2(DHkey, N_{a0}, N_{b0}, r_a)$ at device A and $r=f2(DHkey, N_{a0}, N_{b0}, r_b)$ at device B 
 and set the six most significant digits of r to r$^*_a$. This protocol needs a generation of two additional 128-bit random nonces as well as two additional cryptographic hash function f2() which is originally used in the generation of link key in SSP phase 4.

\begin{figure}[htbp]
\centerline{\includegraphics[width=0.6\linewidth]{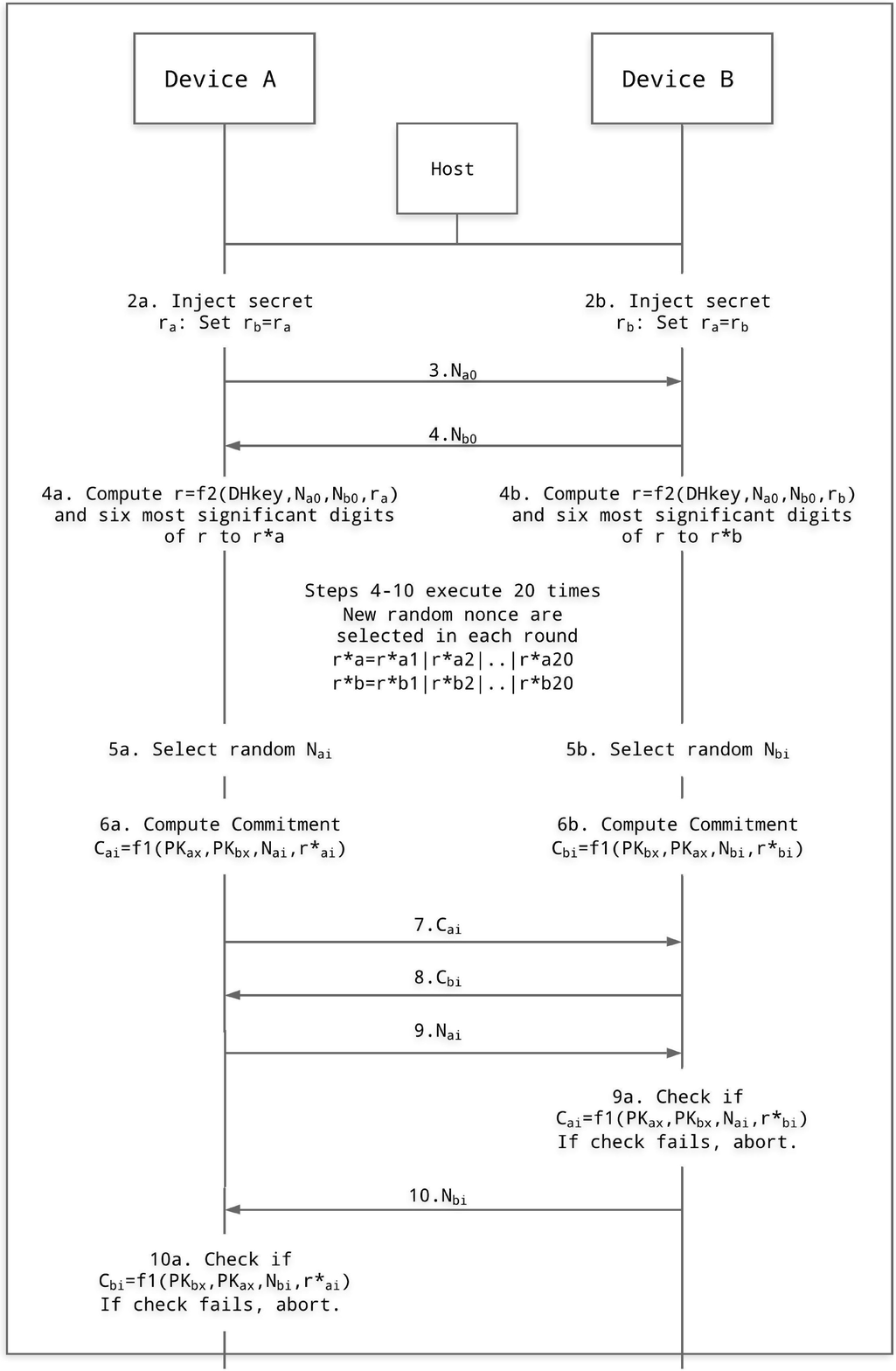}}
\caption{SM Improved Passkey Entry Model.}
\label{fig:sm}
\end{figure}

 in Figure~\ref{fig:sm}, after injecting the passkey r on both sides, the devices A and B generate two random nonces Na0 and Nb0. The two nonces are sent to each other. Then, on device A, a new hash value is computed as $r=f2(DHkey, N_{a0}, N_{b0}, r_a)$ and the six most significant digits of r is set to \textbf{r$\boldsymbol{^*_a}$}. The device B computes the hash value $r=f2(DHkey, N_{a0}, N_{b0}, r_b)$ and sets the six most significant digits of r to \textbf{r$\boldsymbol{^*_b}$}. The rest of the protocol, the steps 5-10, is run the same as that of the original passkey entry model. In their protocol, the random passkey r is used as a seed of the authentication passkey.

This way, even if the user uses the same passkey next time, the attacker cannot easily attack since each time the values of \textbf{r$\boldsymbol{^*_a}$} and \textbf{r$\boldsymbol{^*_b}$} will be different.

\subsection{An attack on SM protocol}
Consider a scenario where an attacker is simultaneously trying to establish a connection with both devices A and B. If we assume that the device A and device B have a fault tolerance system where they allow three to four consecutive SSP sessions with minimal delay in the event of failure, the attacker can try to establish consecutive sessions to device A and device B event in the event of failure of prediction of bits. When an attacker has successfully completed the public key exchange, then they have two DHKey values, one with device A and another with device B. If we assume that an attacker was able to correctly predict around 8-bits of the r$^{*}_a$ and r$^{*}_b$, then a simple brute-force would greatly increase the chances of the attacker to predict the correct passkey. This is because aside from the passkey r, the attacker knows the values of DHkey, N$_{a0}$, and N$_{b0}$. So after the SSP session fails due to the incorrect passkey bit, the attacker can simply brute-force all the combinations of passkey and match the output with the 8-bits of r$^{*}_a$. Out of the passkeys which match, there is going to be the correct passkey. The attacker can now take the list of the matched passkeys and perform a dictionary attack and try those combinations with the first 8 bits of r$^{*}_b$ of another SSP session.  

The algorithm is displayed in Algorithm \ref{alg1}.

\begin{algorithm}[htbp]
 \caption{Brute-force algorithm}\label{alg1}
  \begin{algorithmic}[1]
   \State Initialise DHkey, N$_{a0}$, N$_{b0}$, newlist[] and count = 0 obtained from first SSP session.
   \For{i in range of (0..1000000)}
    \If {i$\leq$100000}
        \State prepend zeros to i to make it 6-digit.
    \EndIf
    \State Generate r$_x$ = HMAC-SHA256(DHkey, N$_{a0}$, N$_{b0}$, i)
    \State Set n most significant of r$_x$ to r$'_x$ where n is the obtained number of r$^{*}_a$ bits by the attacker.
    \If {r$'_x==$r$^{*}_a$}
        \State newlist[count] = i (Adding i to the list of potential passkeys.)
        \State count = count + 1
        \State print(Match found. Potential Passkey = i)
    \EndIf
   \EndFor
 \Procedure{consecutivebrute}{DHkey, N$_{a0}$, N$_{b0}$, newlist, count}
    \State Initialise DHkey, N$_{a0}$, N$_{b0}$ obtained from next consecutive SSP session.
    \State newcount = 0
    \For{i in range of (0..count)}
    \If {newlist[i]$\leq$100000}
        \State prepend zeros to newlist[i] to make it 6-digit.
    \EndIf
    \State Generate r$_x$ = HMAC-SHA256(DHkey, N$_{a0}$, N$_{b0}$, newlist[i])
    \State Set n most significant of r$_x$ to r$'_x$ where n is obtained number of r$^{*}_a$ bits by the attacker.
    \If {r$'_x==$r$^{*}_a$}
        \State newlist[newcount] = i (Adding i to the list of potential passkeys.)
        \State newcount = newcount + 1
        \State print(Match found. Potential Passkey = newlist[i])
    \EndIf
   \EndFor
\EndProcedure
\State Execute consecutivebrute(DHkey, N$_{a0}$, N$_{b0}$, newlist, count) with a new list of potential passkeys until the correct passkey is found.
 \end{algorithmic}
\end{algorithm}

By doing this, the attacker can eventually obtain the correct passkey within 2-3 SSP sessions. When the passkey is known, the attacker will simply execute  MITM attack in the new session,  and this time, they will be successful because of the correct passkey and will be able to obtain the LTK.

\subsection{Simulation Results}
\begin{figure}[htbp]
\centerline{\includegraphics[width=0.5\linewidth]{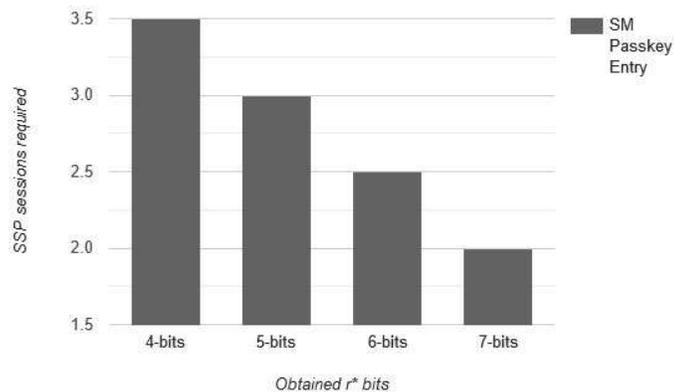}}
\caption{SM protocol resistance against Brute-force attack.}
\label{fig:bar}
\end{figure}

We have done a simulation where we tested the resilience of the SM passkey entry model against the brute-force attack. For this specific simulation of the brute-force attack, we used a desktop computer with an Intel i7 2.20GHz processor. We have performed the brute-force attack when 4, 5, 6 and 7 bits of r* are known to the attacker. The brute-force attack was conducted 50 times with each known r* value and the average SSP sessions required were calculated for each r* value as shown in Figure~\ref{fig:bar}.

\section{Enhanced Passkey Entry Protocol}
To improve the security of the passkey protocol, we propose an alternate to \cite{dazhimitm}, our enhanced passkey entry protocol below.
Our protocol does not differ much from the original protocol in Bluetooth standard and in addition to that, there are only 10 rounds in our protocol which significantly reduce the power and communication costs of the devices. The main idea of our protocol is to never reveal the original passkey until phase 3 of the SSP. We also make use of the properties of the DHKey in the ECDH algorithm to generate a 256-bit hash from passkey r. Therefore, we use a derivative hash of the original passkey r and use that to run the entire passkey entry protocol. In our protocol, there are 2 different bits exchanged and verified at each side for each round resulting in 20-bits being exchanged altogether.

\begin{figure}[htbp]
\centerline{\includegraphics[width=0.6\linewidth]{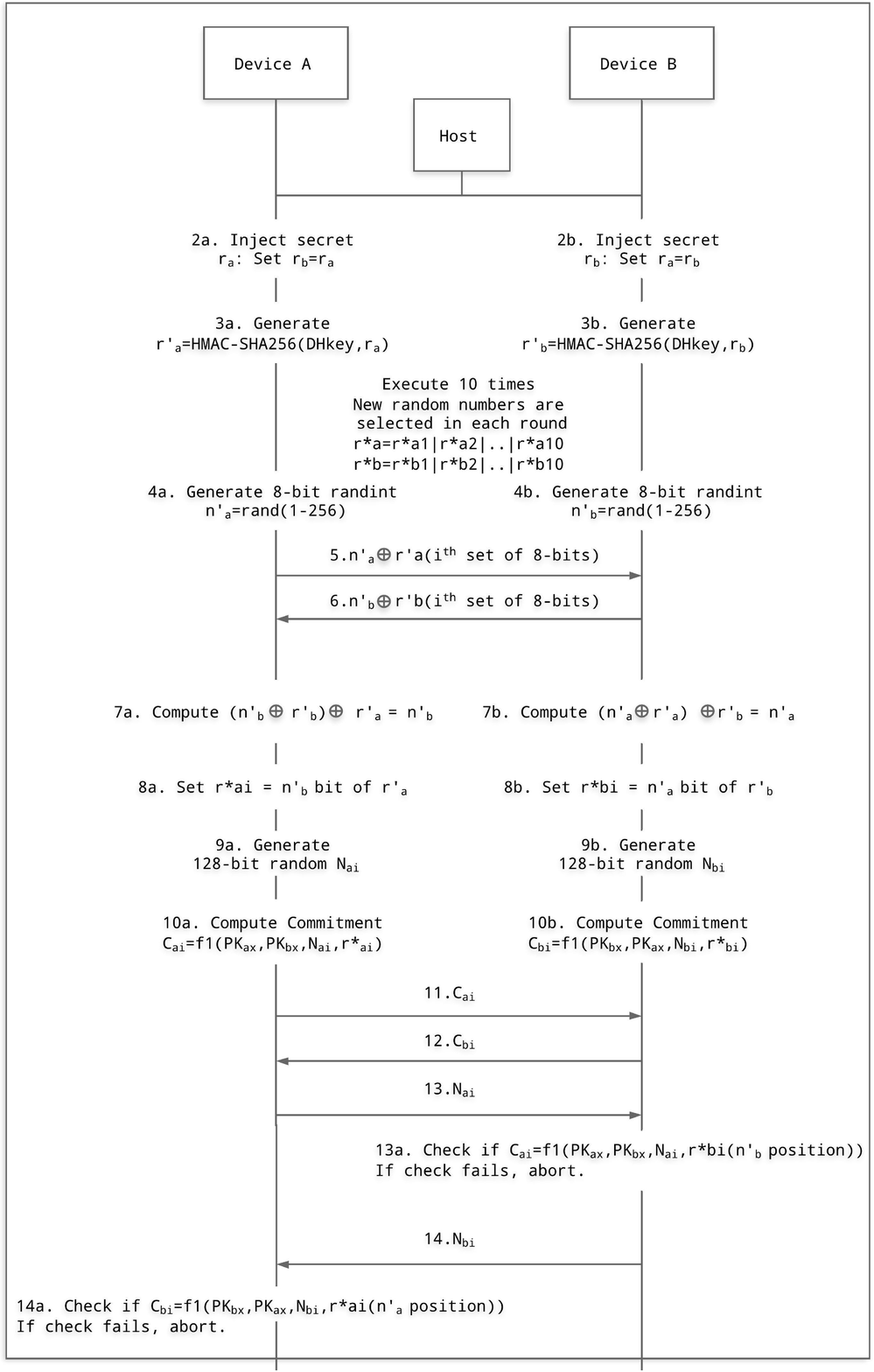}}
\caption{New Enhanced Passkey Entry Protocol.}
\label{Fig:Enhanced}
\end{figure}

Our protocol is displayed in Fig.~\ref{Fig:Enhanced}. When the user enters the random passkey on both devices, we make use of the f2() hash function of the original passkey entry protocol to create a 256-bit hash value from the input passkey r. We use the shared DHKey calculated in the public key exchange phase as the key for the HMAC function and the input message, 20-bit passkey. The resultant hash is denoted as \textbf{r$\boldsymbol{'_a}$}  and \textbf{r$\boldsymbol{'_b}$} respectively. The computed hashes \textbf{r$\boldsymbol{'_a}$} and \textbf{r$\boldsymbol{'_b}$} calculated in steps 3a and 3b have the same value since the passkey is same at both devices. After generating these values, the steps 4-12 are executed for 10 times. In steps 4a and 4b, an 8-bit random integer \textbf{n$\boldsymbol{'}$} (where $1<=n\boldsymbol{'}<=255$) is generated at both sides i.e., \textbf{n$\boldsymbol{'_a}$} and \textbf{n$\boldsymbol{'_b}$} for device A and device B respectively. This random integer defines the position at which the passkey bit is to be taken from \textbf{r$\boldsymbol{'_a}$} and \textbf{r$\boldsymbol{'_b}$}. After generating the random integers, both devices perform an XOR operation with the 8-bit random integer \textbf{n$'$} and the $\boldsymbol{i^{th}}$ set of 8-bits of the \textbf{r$\boldsymbol{'}$}. Both values are exchanged in steps 5 and 6. Next, at device A, we compute the XOR as \textbf{$(n\boldsymbol{'_b} \oplus r\boldsymbol{'_b}) \oplus r\boldsymbol{'_a}$} and obtain the \textbf{n$\boldsymbol{'_b}$} value. Then, we set the \textbf{n$\boldsymbol{'_b}$}th bit of 256-bit \textbf{r$\boldsymbol{'_a}$} hash value to the \textbf{r$\boldsymbol{^*_{ai}}$} bit. For instance, if the random number \textbf{n$\boldsymbol{'_b}$} is 20, then we set the value of the 20$^{th}$ bit of \textbf{r$\boldsymbol{'_a}$} hash as \textbf{r$\boldsymbol{^*_{ai}}$}. The same process is followed at the device B where we calculate \textbf{$(n\boldsymbol{'_a} \oplus r\boldsymbol{'_a}) \oplus r\boldsymbol{'_b}$} to obtain the \textbf{n$\boldsymbol{'_a}$}. Since the values of the \textbf{r$\boldsymbol{'_a}$} and \textbf{r$\boldsymbol{'_b}$} are equal, the XOR operation will derive the correct values. After setting the \textbf{r$\boldsymbol{^*_{ai}}$} and \textbf{r$\boldsymbol{^*_{bi}}$} values, both devices generate a 128-bit random nonce. In steps 10a and 10b, the device A computes a commitment value C$\boldsymbol{_{ai}}$$=f1$(PK$_{ax}$,PK$_{bx}$,N$_{ai}$,r$^{*}_{ai}$) and device B computes a commitment value C$\boldsymbol{_{bi}}$$=f1$(PK$_{bx}$,PK$_{ax}$,N$_{bi}$,r$^{*}_{bi}$). Both commitment values are exchanged between the two devices. Then, the device A sends the 128-bit random nonce \textbf{N$\boldsymbol{_{ai}}$}. After receiving the random nonce \textbf{N$\boldsymbol{_{ai}}$}, the device B sets the \textbf{r$\boldsymbol{^*_{bi}}$} with the value at \textbf{n$\boldsymbol{'_b}$}$^{th}$ bit of \textbf{r$\boldsymbol{'_b}$} hash. Next, the commitment value \textbf{C$\boldsymbol{_{ai}}$} is checked by verifying C$\boldsymbol{_{ai}}$$=f1$(PK$_{ax}$,PK$_{bx}$,N$_{ai}$,r$^{*}_{bi}(n'_b position)$). If the two  values match, then the protocol is continued. The device B sends the random nonce \textbf{N$\boldsymbol{_{bi}}$} and random integer \textbf{n$\boldsymbol{'_b}$}. The same process is followed at device A and the commitment values are verified. This process is repeated for 10 rounds where two bits of the 256-bit \textbf{r$\boldsymbol{'}$} hash are sent randomly by two devices in each round.

The main idea of our proposed protocol is to use bits from the random position of the hash value of the passkey to do the commitment instead of using the bits of passkey. To keep the position information secure, we again use the hash value as the one-time pad encryption key to encrypt the information of positions. In our protocol, the passkey is never used to communicate between two parties.Therefore no information will be revealed by the communication.
 Since the random position is used, Algorithm~\ref{alg1} does not work for our protocol.

\section{Security of Enhanced Passkey Entry Protocol under passive Eavesdropping and MITM attacks}

Compared to the original passkey entry protocol, our enhanced protocol provides protection against passive and active attacks. Consider a scenario where the attacker X captures all the commitment values and the random nonce of the passkey entry protocol run using passive eavesdropping. Note that here, we assume that the attacker already has the knowledge of the frequency hopping pattern of the target devices. By doing this, the attacker effectively obtains the public keys of device A and device B. Then in the passkey entry protocol run, the commitment values and the random nonces are obtained. Now, the attacker X can try to deduce the value of \textbf{r$\boldsymbol{^*}$} bits by comparing the commitment values but X cannot find out the original passkey r from the \textbf{r$\boldsymbol{^*}$} bits. Therefore, even though X knows \textbf{r$\boldsymbol{^{*}}$}, they cannot deduce the original passkey  because the \textbf{r$\boldsymbol{^*}$} bits are generated from the hash function which uses the original passkey r as the message and the DHkey as the key for the HMAC operation. In a worst case, the attacker would have 10 bits of the \textbf{r$\boldsymbol{^*_{ai}}$} and \textbf{r$\boldsymbol{^*_{bi}}$} values which are exchanged in each round. However, these values are not enough to deduce the entire hash. In addition to this, even if the DHkey is known  the attacker would still not know the position of the \textbf{r$\boldsymbol{^{*}}$} bits which are exchanged due to XOR operations. In addition to this, the \textbf{r$\boldsymbol{'}$} value will be different for every SSP session which leads to an entirely different and unique hash value \textbf{r$\boldsymbol{'}$} in another SSP session even a same passkey is used. In this way we eliminate the possibility of an attacker that uses passive eavesdropping to deduce the original passkey. To summarise, it would be impossible for an attacker to deduce the full hash value of the passkey even if they passively eavesdropped the target device's SSP sessions continuously several times because of the following reasons.

\begin{itemize}
\item Even if the passkey is reused, the hash value obtained in step 3 of Figure~\ref{Fig:Enhanced} will be different for every SSP session. Each time, the target device connects to a different device, the DHKey value will change and as a result, the hash value will also change.
\item The ECDH private-public key pair is generated randomly and is not static.
\item Since the n' bits are hidden using the XOR operations, an adversary cannot guess the correct passkey due to a one-time pad technique.
\end{itemize}

We have also compared the security of our protocol to the SM protocol. While the SM protocol can be attacked using our brute-force attack mentioned in algorithm~\ref{alg1}, the enhanced passkey entry model completely prevents an adversary from establishing the attack. This is because of the XOR operations done on both sides to hide the correct value of the \textbf{n$\boldsymbol{'_a}$} and \textbf{n$\boldsymbol{'_b}$}. From an attacker's perspective, the XOR'ed value which is sent during steps 5 and 6 is not useful to them. Unless the attacker already knows the random integer \textbf{n$\boldsymbol{'}$} or the $i^{th}$ set of 8-bits of the correct hash value, they can obtain neither the correct \textbf{n$\boldsymbol{'}$} or the $i^{th}$ set of \textbf{r$\boldsymbol{'}$} value. In the XOR operation, the value of the \textbf{n$\boldsymbol{'}$} and the set of 8-bit \textbf{r$\boldsymbol{'}$} used is different for each round and is not repeated anywhere. Therefore, the XOR operation that is executed in steps 5 and 6 technically uses the properties of the one-time pad technique making it very secure.

In case of a MITM attack, when the public key exchange phase is completed between the attacker and the devices A and B. Now, the attacker will share one DHkey with device A and one DHkey with device B. In our proposed protocol, after the legitimate user enters the 6-digit passkey, we generate a hash using the DHkey and the 6-digit passkey. Since the attacker does not know the passkey, they have 1 million possibilities of the passkey. Even if one digit of the passkey is guessed wrong, the hash value which is computed is changed drastically. In addition to this, the hash value of device A and device B will be different due to the different DHkey values. Due to this, when the XOR operation is done with the 8-bit random integer and the 8-bit set of \textbf{r$\boldsymbol{'}$} and the result is exchanged between the devices, the derived value will be different leading to an incorrect \textbf{n$\boldsymbol{'}$} value. Therefore, even if an attacker predicts a correct \textbf{r$\boldsymbol{^{*}}$} bit in one round, they would not know the correct value of the position \textbf{n$\boldsymbol{'}$} which is needed to execute the brute-force attack described in Algorithm~\ref{alg1}.

Therefore, our protocol not only thwarts off passive eavesdropping but also Man-in-the-middle attacks even when a passkey is reused.

\section{Performance Evaluation}
\subsection{Performance Cost of Original Passkey Entry Model}
In the original passkey entry protocol for each round, there are hash operations of f1 at steps 4a, 4b, 7a, and 8a as shown in Figure~\ref{Fig:SSP2}. Therefore, the computation cost between the two devices is 80 hash functions. Coming to the commitment values, each value has a size of 128bits. And there are exactly 2 commitment values exchanged in each round for 20 rounds. Therefore, the communication cost for the commitment values can be calculated as 

$$20(rounds) * 2(commitment values) * 128(sizeof(commitment)) = 5120bits$$
 And in addition to that, the two devices exchange exactly two random nonce for each round in $1\le i \le 20$ which leads to 
 
 $$20(rounds) * 2(Random Nonce) * 128(sizeof(randnonce)) = 5120bits$$
  Altogether, we have a total of 10,240 bits being exchanged in the protocol run. Finally, the storage cost of the protocol would be around 192 or 256 bits depending on the length of the DHkey. Taking the size of the passkey into account, we also need an additional 20-bits. Then, the total storage cost would be 212 or 276 bits.

\subsection{Performance Cost of Enhanced Passkey Entry Model}
Our enhanced protocol performs in the same way as the original. The main difference of our protocol is that it significantly reduces the number of hash computations required due to the 10 rounds. We also need two extra variables to store the calculated 256-bit hash $\boldsymbol{r'}$ and the 8-bit random integer $\boldsymbol{n'}$. In addition to this, there are 2 additional hash computations present before the execution of steps 4-14 in Figure~\ref{Fig:Enhanced}. These hash computations are done on both sides. We are not considering the XOR operations done on both sides as the operation cost is almost negligible. In our protocol, aside from two hash computations in step 3a and 3b, in each round, there are 4 hash computations which are used to compute the commitment values and verify them at both sides. So, for 10 rounds we have 4(Hash) x 10(rounds) = 40 Hash computations. Therefore, the computation cost of our protocol adds up to a total of 42 Hash computations. Regarding the communication cost, our protocol follows the same steps as that of the original protocol aside from the fact that two 8-bit random integers are exchanged in each round.  The size of the commitment values and the random nonces remain the same. Therefore, the communication cost can be calculated as 

\

 $10(rounds) * 2(randint) * 8(randint size) + 10(rounds) * 2(commitment values) * 128(commitment size) + 10(rounds) * 2(random nonces) * 128(random nonce size) = 5280bits$

\

The storage cost of our protocol is 468 or 532 bits (including the 20-bit passkey) depending on the length of DHkey. Although the storage cost is higher than the original protocol, it significantly increases the security of the protocol by providing two layers of security and protecting devices against passive attacks and active MITM attacks.

Our enhanced passkey entry significantly reduces the computation cost as well as the communication cost in half. Due to the decrease in the computation cost, the device will not take as much power as the original protocol. In addition to this, the passkey entry protocol runs faster due to the decrease in the communication cost as we only need to send half of the bits compared to the original protocol.

\section{Conclusion}
In this paper, we have introduced a new enhanced passkey entry protocol that successfully protects the Bluetooth devices against passive attacks and MITM attacks while reducing the computation and communication costs by half. It should be noted that we have designed our protocol without deviating too much from the original protocol. This is done so that our enhanced protocol can be simply integrated with the Bluetooth devices in the form of a patch. We have made a theoretical approach to the passkey reuse issue and protection against MITM in the passkey entry model. We will  consider employing our proposed protocol in different Bluetooth devices and test the protocol under different scenarios as our future work.

\end{document}